\documentclass{article}
\newcommand{\bb}{\begin{eqnarray}}
\newcommand{\ee}{\end{eqnarray}}
\begin{document}
\title{\bf Time-reversal and parity conservation for gravitating quarks}
\author{P. Mitra\\
Saha Institute of Nuclear Physics\\
Block AF, Bidhannagar\\
Calcutta 700064}
\date{}
\maketitle
\begin{abstract}
The complex mass term of a quark does not 
violate time-reversal or parity in gravitational interactions,
in spite of an axial anomaly.
\end{abstract}

\bigskip

Time-reversal is known to be violated in the weak interactions.
Mass terms $\bar q_L {M} q_R+\bar{\tilde q}_L {\tilde M}\tilde q_R$+hc are generated
for quarks $q$ having charge $+\frac23$ and quarks $\tilde q$ 
having charge $-\frac13$ in the shape of complex matrices $M,\tilde M$ 
from electroweak symmetry breaking.
On diagonalization of ${M,\tilde M}$ through flavour matrices,
\bb
q_L\rightarrow A_L^{-1}q_L,&&q_R\rightarrow A_R^{-1}q_R\\
\tilde q_L\rightarrow  \tilde A_L^{-1}\tilde q_L,
&&\tilde q_R\rightarrow  \tilde A_R^{-1}\tilde q_R,
\ee
charged current weak interactions pick up a matrix $A_L \tilde A_L^{-1}\equiv {C}$, 
the Cabibbo-Kobayashi-Maskawa matrix, 
which is complex and violates time-reversal. 
The diagonalized mass terms too may be complex.
One can write 
\bb
\bar\psi_L me^{i\theta}\psi_R+hc&=&\bar\psi me^{i\theta\gamma_5}\psi\nonumber\\
&=&\cos\theta\bar\psi m\psi+i\sin\theta\bar\psi m\gamma_5\psi.
\ee
This looks like a combination of a scalar and a pseudoscalar,
suggesting parity violation. More important,  
the phase factor $e^{i\theta}\rightarrow e^{-i\theta}$  under
antilinear operations,
suggesting additional time-reversal violation
in the standard model beyond the weak interaction violation
mentioned above.

A chiral transformation may be used to remove $\theta$:
\bb
\psi&\rightarrow& e^{-i\theta\gamma_5/2}\psi,\nonumber\\
\bar\psi&\rightarrow& \bar\psi e^{-i\theta\gamma_5/2}.
\ee
$\theta$ does get removed from the mass term, but may reappear
as a topological term in the gluon sector \cite{baluni} through
the Jacobian of the fermion measure which is not invariant under
the chiral transformation. This is because of the axial anomaly.
However, the question of time-reversal
violation depends more subtly on the fermion measure and is avoided in
quantum chromodynamics with an appropriate choice of measure \cite{bcm}.

What happens if one considers the gravitational
coupling of the quarks? 
In a gravitational field too, there is an axial anomaly 
\cite{anom} given by
\bb
{\epsilon^{\mu\nu\alpha\beta}R_{\mu\nu\sigma\tau}R_{\alpha\beta}{}^{\sigma\tau}
\over 384\pi^2}.
\ee
Time-reversal and parity violations could show up
in gravitational interactions of quarks in the same way that strong
interactions had been thought to involve these symmetry violations.
This is the question we analyze here.

If there is a complex fermion mass term in curved spacetime, its phase $\theta$ 
may again be removed from the fermion action by a chiral transformation, 
so that there is no
direct parity or time-reversal violation,
suggesting a redefinition of parity 
and time-reversal at least at the {\it classical} level.
Whether there is any indirect
violation of the symmetries through $R\tilde R$ terms generated by the
chiral transformation will be investigated below through
the fermion determinant.

The Dirac operator without any mass term may be written as
\bb
\not D=\gamma^le^\mu_l(\partial_\mu-iA_\mu-\frac{i}{2}A^{mn}_\mu\sigma_{mn}).
\ee
Here, $e^\mu_l$ is the  tetrad, $A^{mn}_\mu$ is the spin connection,
and $\sigma_{mn}\equiv i[\gamma_m,\gamma_n]$. $A_\mu$ is any other 
gauge field coupling vectorially to the fermion.

For a real mass term, the fermion action is
\bb
\int\bar\psi (i\not D -m)\psi.
\ee
This is invariant under the standard parity transformation
\bb
\psi(\vec x)&\to& \gamma^0\psi(-\vec x),\\
e^\mu_l(\vec x)&\to& \pm e^\mu_l(-\vec x),
\ee
where the $\pm$ sign is negative if an odd number of
Greek or Latin spatial indices is involved.
$A^{mn}_\mu$ also changes, but it is determined by $e^\mu_l$.
Note that the invariance involves more than the usual
flat spacetime symmetry: the $\sigma_{mn}$ matrices produce $\pm$ signs
when $\gamma^0$ is taken across it, but these exactly cancel
with the $\pm$ signs from any Latin spatial indices of $A^{mn}_\mu$.

Similarly, the action is also invariant under the time-reversal transformation
\bb
\psi(x^0)&\to& i\gamma^1\gamma^3\psi(-x^0),\\
e^\mu_l(x^0)&\to& \pm e^\mu_l(-x^0),
\ee
where the $\pm$ sign is negative if an odd number of
Greek or Latin temporal indices is involved. 
Once again the $\sigma_{mn}$ matrices produce $\pm$ signs
which cancel $\pm$ signs from Latin temporal indices of $A^{mn}_\mu$.

The fermion action with a {\it complex} mass term is
\bb
\int\bar \psi(i\not D -m\exp(i\theta\gamma^5))\psi.
\ee
The mass term involves the matrix $\gamma^0\exp(i\theta\gamma^5)$
and it is not difficult to guess that the action will be
invariant under the chirally rotated parity transformation
\bb
\psi(\vec x)\to \gamma^0\exp(i\theta\gamma^5)\psi(-\vec x).
\label{p}\ee
The chiral phase factor $\exp(i\theta\gamma^5)$ commutes with $\sigma_{mn}$,
which does not therefore provide any additional complication.

Like the parity transformation, the action is also
invariant under a chirally rotated time-reversal transformation
\bb
\psi(x^0)\to i\exp(i\theta\gamma^5)\gamma^1\gamma^3\psi(-x^0).
\label{t}\ee

These discrete symmetries have been seen so far at the classical level.
The symmetries
could break down at the quantum level, {\it i.e.,} they could be
anomalous. There is in fact a reason to suspect such an eventuality
because the symmetry transformations involve chiral rotations,
and chiral transformations are known to develop anomalies upon quantization. The
question here is whether there exist ways of regularizing the
theory so that the classical {\it discrete} symmetries are preserved.

Let us use a zeta function regularization \cite{zeta}. 
The Dirac operator
\bb
\not D=\gamma^le^\mu_l(\partial_\mu-iA_\mu-\frac{i}{2}A^{mn}_\mu\sigma_{mn})
\ee
is hermitian for euclidean signature in the scalar product
\bb
(\phi,\psi)\equiv\int d^4x\sqrt{g}\phi^\dagger\psi,
\ee
but its combination with a mass term, real or complex,
is not hermitian. A positive operator, needed for defining
the zeta function, can then be defined as
\bb
\Delta&=&[i\not D-m\exp (i\theta\gamma^5)]^\dagger\nonumber\\
&&[i\not D-m\exp (i\theta\gamma^5)].
\ee
It is easy to see that
\bb
\Delta=(\not D)^2+m^2.
\ee
Hence $\theta$ does not enter $\Delta$ or the zeta function 
\bb
\zeta(s,\Delta)\equiv{\rm Tr}(\Delta^{-s}).
\ee
The logarithm of the fermion determinant, defined as
\bb
-\frac12\zeta'(0,\Delta)-\frac12\ln\mu^2\zeta(0,\Delta),
\ee
with some $\mu$, is also independent of $\theta$.
Since there is no violation of parity or time-reversal at $\theta=0,$
it follows that there is no violation even for non-zero values. In
other words, the parity and time-reversal symmetries found above for
the classical action involving quarks with a complex fermion mass term
are not anomalous and survive quantization even in curved spacetime.
The $\gamma^5$ phase $\theta$ violates neither parity nor time-reversal
even in gravitational interactions, just as in the
strong interactions of quarks \cite{pm}.
As in that case, violations can occur if there are 
direct $R \tilde R$ terms \cite{ddi}.

Some remarks about anomalies may be apposite here. If a fermion is
coupled to a gauge field or a gravitational field, the fermion
measure has to be defined with care and is not invariant
under a chiral transformation of the fermion. This is the meaning
of the statement that there is a chiral or axial anomaly. However,
what we are considering here are time-reversal and parity
transformations. We need to know whether the measure is invariant
under these discrete transformations. A possibility of non-invariance
arises because the discrete transformations (\ref{p},\ref{t})
involve chiral transformations. If the measure is not invariant,
the classical symmetries will be anomalous and the fermion
determinant will not have the symmetries. 
The construction of the fermion
determinant directly shows that it is independent of $\theta$
and therefore has time-reversal and parity symmetries. This means
that the measure implicit in the zeta function regularization
is invariant under (\ref{p},\ref{t}) and these transformations are
free from anomalies. The classical fermion action also has
gauge invariance, general coordinate invariance and local Lorentz
invariance. These are all maintained by the measure of Dirac
fermions.

\end{document}